\documentclass{nature}
\usepackage{graphics,graphicx}
\usepackage{url}
\usepackage{enumitem}
\usepackage{color}
\usepackage{placeins}

\newcommand\old[1]{}

\title{MetaScope - Fast and accurate identification of microbes in metagenomic sequencing data}

\author{ 
Benjamin Buchfink$^{1}$,
Daniel H. Huson$^{1,2}$
\& Chao Xie$^{2,3}$
}

\begin{document}
\spacing{1}

\maketitle

\begin{affiliations}
 \item Department of Computer Science and Center for Bioinformatics, University of T\"ubingen, Sand 14, 72076 T\"ubingen, Germany
\item Life Sciences Institute, National University of Singapore, \#05-02 28, Medical Drive Singapore 117456 Singapore
\item Human Longevity Incorp., Singapore
\end{affiliations}

\begin{abstract}
MetaScope is a fast and accurate tool for analyzing (host-associated) metagenome datasets.
Sequence alignment of reads against the host genome (if requested) and against
microbial Genbank is performed using a new DNA aligner called SASS. The output of SASS is
processed so as to assign all microbial reads to taxa and genes, using a new weighted version of the LCA algorithm.
MetaScope is the winner of the 2013 DTRA software challenge entitled ``Identify Organisms from a Stream of DNA Sequences''. 

\end{abstract}

Metagenomics is the study of microbes using  DNA sequencing\cite{Handelsman2004}.
One major area of application is the human microbiome\cite{HumanMicrobiomeProject2007} with the aim of understanding
the interplay between human-associated microbes and disease. Other areas of metagenomic research include
water\cite{Rusch2007}, waste-water treatment\cite{Nielsen2011}, soil\cite{Mackelprang2011} and ancient  pathogens\cite{Schuenemann2011}. Another envisioned  area of application metagenomics is in bio-threat detection, for example,
when a group of individuals becomes infected by an unknown agent  and the goal is
to  quickly and reliably  determine the identity of the pathogens involved.

In 2013
the Defense Threat Reduction Agency (DTRA) sponsored an algorithms competition entitled ``Identify Organisms from a Stream of DNA Sequences'' with a one million dollar prize.
Proposed solutions ``must generate equivalent identification and characterization performance regardless of the sequencing technology used'' and ``\dots must achieve this in a timeline that is substantively shorter than possible with currently available techniques.'' The challenge provided nine test datasets  for analysis and proposed results 
were scored based on the correctness of organisms identified (organisms score), reads assigned (reads score) and
 genes identified (genes score).  This paper describes the winning entry.

Such analysis requires the comparison of a large number of sequencing reads (typically millions of reads) against a large reference database (typically many billions of nucleotides or amino acids). Hence, tools that address this type of problem must
be very fast. Because current reference databases only represent a small fraction of the sequence diversity that exists in the environment\cite{GEBA2009}, such tools must also be very sensitive.

MetaScope performs very fast and very accurate analysis
of metagenome datasets, including the removal of host reads, if required. MetaScope employs  a new fast and sensitive
DNA aligner called SASS. The aligner is first used to compare a given set of metagenomic sequencing reads
 against a host genome, if available, so as to discard reads that probably come from the host genome. The remaining reads
 are then compared against microbial Genbank\cite{Wheeler2008} using SASS. 
 A second program called Analyzer processes the output of SASS and maps the reads
to taxa and genes using a novel variant of the LCA algorithm\cite{MEGAN2007}.  The output is written in XML and can, for example,
 be loaded into  the metagenome analysis program MEGAN\cite{MEGAN2011} for further processing. 

Like BLAST\cite{altschul90}, SASS uses a seed-and-extend approach to alignment. To achieve both high speed and high accuracy, SASS uses
spaced seeds\cite{Burkhardt01,Ma02}, a hash-table for seed lookup and is implemented as a parallel algorithm in C++\cite{SeqAn}. 
A crucial heuristic decision is when to extend a given seed match so as to compute a full alignment.
SASS uses Myers' bit vector algorithm\cite{Myers1999b} and a gain-based termination criterion to decide this. In the context of host-genome filtering, the score obtained in this way is used as a proxy for
the full local alignment score and the extension phase is not used.

Removal of all reads that align to the host genome does not completely resolve the problem of false positive taxon assignments because many viruses and vector sequences in Genbank  contain human sequences. Hence, in a preprocessing step, we use SASS to compare the
viral and vector portion of  Genbank  against the human genome and then mask every region of the reference sequences  that has a significant alignment to some host sequence.

The assignment of reads is often performed using the naive
LCA algorithm\cite{MEGAN2007} in which a read is placed on the lowest-common ancestor  of all
taxa in the NCBI taxonomy for which the read has a high-scoring alignment to a corresponding sequence in the reference database.
As the naive LCA algorithm analyses each read in isolation, in the presence of many similar
reference sequences from different species, the result is often very unspecific placement of reads.
 To overcome this, MetaScope uses a new weighted LCA  algorithm that proceeds in two steps. First,
 the naive LCA algorithm is used
to assign a weight to each reference sequence, reflecting  the number of reads that are assigned to the corresponding species and have a significant alignment to that reference sequence.
Then each read is placed on the taxon node that covers 75\% (by default) of the total weight
of all reference sequences that have a significant alignment with the read.

MetaScope  predicts genes based  on which annotated genes the alignments of a read overlap with.
A read will often align to many different reference sequences and so the potential number of genes to report for a single read can be quite large, containing many false positives. To address this,
all genes that are partially overlapped by some alignment of a read are ranked by the weight of the corresponding
reference sequence and by the proportion of the gene sequence covered by any reads, 
and a small number of top ranked genes are reported.

The results obtained by MetaScope on the nine DTRA  datasets are listed in Table~\ref{results-table}.
Slightly different algorithmic parameters are used based on the different sequencing platforms, as described in the Materials section.
The accuracy score ranges from $90.1-98.7\%$ and the run time ranges from $4$ to $13$ minutes per dataset. 

\begin{table}
{\footnotesize
\begin{tabular}{ll*{8}{c}}
\hfil(1) & \hfil (2) & (3) & (4) & (5) & (6) & (7) & (8) & (9) & (10)\\
\hfil Name &  \hfil Sequencing & Number & Average &	Seq. &Total & Org. & Reads & Genes &	Time\\
          &  \hfil platform & of reads &  length & acc. & score & score & score & score & (mins)\\ \hline

Testing1	& PacBio    	&   \hfill 92\,948	& \hfill 1883	& \hfill 83 	& 90.074	& \hfill 100 	& \hfill 85  	& \hfill 85	& \hfill 7:48\\
Testing2	& PacBio    	&   \hfill 98\,323	&\hfill  1837	& \hfill 83 	& 98.747	&\hfill  100 	& \hfill 98  	& \hfill 98	&\hfill  8:24\\
Testing3	& Ion-Torrent  	&  \hfill 379\,028	&  \hfill 160& \hfill 98 		& 91.949	&\hfill  85  	& \hfill 93  	& \hfill 96	& \hfill 6:28\\
Testing4	& Roche-454  	&  \hfill 399\,671	& \hfill  363& \hfill 99 		& 91.595	& \hfill 75  	&\hfill  99  	& \hfill 99	&\hfill  6:47\\
Testing5	& Illumina 		& \hfill 5\,550\,655	&  \hfill 150& \hfill 100		& 91.538	& \hfill 93  	&\hfill  99  	& \hfill 82	& \hfill 6:14\\
Testing6	& Illumina  	&\hfill  6\,038\,557	&  \hfill 150& \hfill 100		& 95.357	& \hfill 100 	& \hfill 100 	& \hfill 86	&\hfill  7:27\\
Testing7	& Ion-Torrent   	&  \hfill 323\,028	&  \hfill 159& \hfill 98		& 92.258	& \hfill 83  	& \hfill 99  	&\hfill  94	& \hfill 4:20\\
Testing8	& Roche-454  	&  \hfill 351\,799	& \hfill  263& \hfill 99		& 96.843	& \hfill 100 	& \hfill 100 	& \hfill 90	&\hfill  4:49\\
Testing10	& Illumina	 	& \hfill 6\,164\,558	& \hfill 151	& \hfill 100	& 97.803	& \hfill 100 	& \hfill 100 	& \hfill 93	& \hfill 12:10\\
\end{tabular}
}
\caption{Nine DTRA challenge human-associated DNA sequencing
datasets (1--5), percentage scores achieved by MetaScope (6--9), and time required (10). Sequencing accuracy (5) is the average percent identity of read alignments to the human reference genome.
The total score (6) is the average of the three component scores (6--8). 
}\label{results-table}
\end{table}

We have also investigated the use of an intermediate assembly step (except for PacBio reads). In more detail, all reads
that did not align to the host genome (human) were presented to the Newbler assembler\cite{Margulies2005short} as input. The obtained contigs  and all unassembled reads
were aligned against microbial Genbank using SASS. The output of this was then
processed as described above and all assembled reads inherited the taxon assignment of their containing contigs.
The results produced by this approach scored equally high as those reported in Table~\ref{results-table}, but not better, so
we did not pursue this further.

We plan to make Metascope freely available from \url{http://www.metascope.net}.

\section*{Methods}

\subsection{Overview}\label{sec-overveiw}

MetaScope is able to accurately analyze millions of sequencing reads in minutes.
The MetaScope pipeline takes a file of sequencing reads in {FastQ} or FastA format as input and produces a report file in {XML} format as output.
The input file represents an host-associated metagenome sample and the aim of MetaScope is to determine the taxonomic and functional content of
the non-host portion of the sample.
For each organism detected in the input file, the report file contains an estimation of its relative amount, the list of all reads assigned to the organism
and the list of all genes identified for the organism. 

\begin{figure}
 \hfil
 \includegraphics[width=\textwidth]{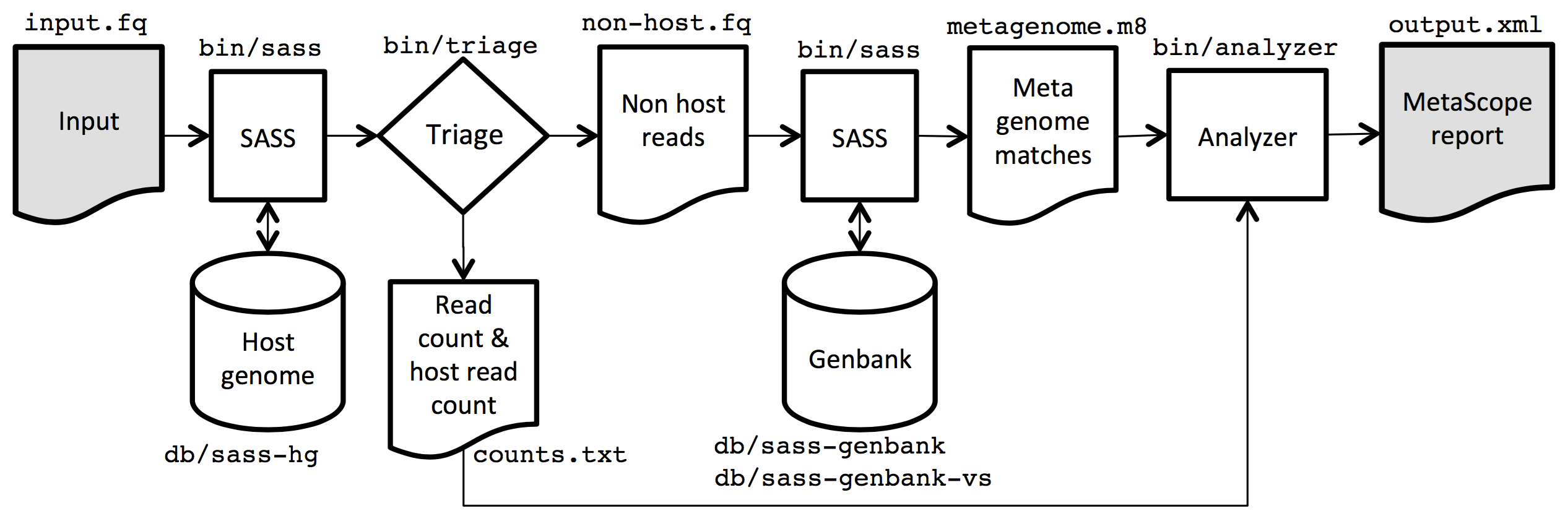}
 \hfil
 \caption{\small Overview of the MetaScope pipeline. Data moves from left to right. The input reads are compared against
the host genome using SASS, then all non-host reads are compared against Genbank using SASS,
and  finally, the result of the second comparison is processed by the MetaScope Analyzer program to generate the
final MetaScope XML report.}\label{pipeline-fig}
 \end{figure}

The MetaScope pipeline (see Figure~\ref{pipeline-fig}) is invoked using  the command
{{\tt metascope}} {\em platform reads work output}.
 The four arguments specify the sequencing platform (one of Illumina, 454, IonTorrent or PacBio), an input file containing all reads in FastQ format, a sample-specific working directory where intermediate files are to be placed,  and the name of the output file.
 In addition, any part of the pipeline can be run individually with more control over programming parameters.
 
First, the program SASS is used to compare all reads in the input file against the {host genome}. All detected alignments
between reads and the host genome are written to a file called {\tt host.m8}.
Second, a script called {\tt triage} uses the alignments in {\tt host.m8} to count the host reads and
to write all non-host reads to a file called {\tt non-host.fq}.
Third, SASS compares the set of all non-host reads against  Genbank. All  found alignments are
written to a file called {\tt metagenome.m8}.  The number of reads and host reads, and
 the file of all metagenome alignments are provided as input to
the MetaScope {\tt analyzer}, which produces the final MetaScope report {\tt output.xml} in XML format.
 
\subsection{Sequence alignment}\label{sec-alignment}

The main computational bottleneck in metagenome analysis is the comparison of the reads against a host database, in the case of a 
host-associated sample,
and then against a comprehensive collection of bacterial and viral sequences, such as Genbank\cite{Wheeler2008}.
To address this problem in an efficient manner, MetaScope introduces a new
sequence alignment tool called {SASS} (an acronym for ``sequence alignment using spaced seeds''). SASS is written in C++ and uses  SeqAn\cite{SeqAn} and  Boost\cite{boost_graph}.

Designed to target significant alignments with a bit score of at least 50, SASS aligns DNA sequencing reads at about 50-100
times the speed of discontiguous MegaBLAST\cite{Ma02}.
Like BLAST, SASS attempts to exhaustively determine all significant alignments, which is crucial for accurate taxonomic analysis.\par

Fast pairwise alignment  programs usually follow the {seed-and-extend} paradigm\cite{altschul90}.
In this two-phase approach, first one searches for exact matches of small parts of the query sequence in the reference database,
 such seed matches are evaluated and those deemed promising
are then followed up in an extend phase that aims at computing a full alignment. 

Existing approaches typically employ an index data structure for the reference database in order to quickly
compute all seed matches between the query sequence and the reference. For example,  Bowtie2\cite{Bowtie2} and BWA\cite{BWA} use a compressed FM-index, which is very memory efficient, but at the expense of a slower access time.
In contrast, SASS uses a hash table, which requires more memory, but is much faster.
The high speed of the index allows SASS to achieve high speed
and good sensitivity even when aligning low quality reads such as produced by PacBio and Ion Torrent sequencers.

\begin{figure}
\hfil
\includegraphics[width=10cm]{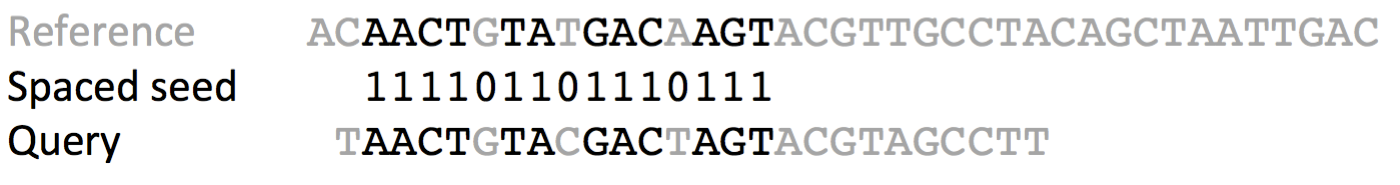}
\hfil
\caption{\small Illustration of the application of a spaced seed to match letters between a reference and a query sequence.  Ones and zeros indicate positions to use and ignore, respectively}
\label{seed-figure}
\end{figure}

Most aligners employ a simple seed shape that consists of a short word of consecutive positions. The choice of seed length is based on a trade-off between sensitivity and speed. 
A hash table index  permits the use of spaced seeds. These are longer seeds in which only a subset of positions are used (see ,  as Figure~\ref{seed-figure}). 
The number and exact layout of the utilized positions are called the weight and shape of the spaced seed, respectively. 
Spaced seeds are known to perform better than simple seeds in terms of the speed/sensitivity trade-off\cite{Burkhardt01,Ma02}.
By default, SASS uses a single spaced seed, {\tt 111110111011110110111111}\cite{Ilie:2011fk}.

To sustain the high throughput achieved in the seeding phase, we attempt to avoid unnecessary  Smith-Waterman computations in the  extension phase. To this end, we evaluate
 seed matches using a modified version of Myers' bit vector algorithm for approximate string matching\cite{Myers1999b}, which computes the edit distance between two short patterns encoded in machine words using fast bit-parallel operations.
 Starting from the location of a seed match, an alignment is extended in both directions by block-wise invocation of Myers' algorithm in conjunction with a termination criterion based on the score gain. 
Tentative scores are calculated that approximate the full BLAST score. 
A full banded Smith-Waterman alignment\cite{LocalAlign} is only performed on the 100 (by default)
best  tentative alignments for a read,
thus producing accurate standardized BLAST alignment scores for them. 

In the case of a host-associated sample, the first step is to identify all reads that come from the host organism.
To address this, SASS is used to compare all reads against the host genome. For human, we used
 assembly release CRCh37 downloaded from NCBI in June 2013. The output of SASS is written to a file called {\tt host.m8}.
To reduce the running time of this calculation, here we compute only the approximated score for any alignment and
do not perform a full Smith-Waterman calculation. 

Based on the host alignments detected in the previous step, a simple Perl script called {\tt triage}
is then used to determine all reads that do not have a significant alignment
to the host genome and only these reads are considered in the downstream analysis. 
Here, an alignment is considered significant if  it has an expected score of less than $10^{-10}$.
These reads are placed in a file called {\tt non-host.fq}.
An additional file, {\tt counts.txt} is generated that contain the total number of reads and the number of reads
that have a significant alignment to the host genome.

Then SASS is used to compare all non-host reads reads (contained in {\tt non-host.fq}) against a large  portion of Genbank (consisting of all bacterial, viral, phage and synthetic sequences), downloaded from
NCBI in June 2013.
The resulting alignments are placed in a file called {\tt metagenome.m8}.

SASS uses two different indices for Genbank, depending on the quality and quantity of  the sequencing reads.
For high quality and high quantity input samples SASS uses an index  that is optimized for speed (using longer seeds) whereas for
samples of lesser sequencing quality and smaller size, SASS uses an index that is optimized for sensitivity (using shorter seeds).


The  files  {\tt counts.txt}  and {\tt metagenome.m8} form the basis of MetaScope's taxonomy and gene content analysis,
as described in the following sections.

\subsection{Taxonomic analysis}

The number of reads and host reads, and
 the file of all metagenome alignments obtained using SASS are provided as input to
the MetaScope {\tt analyzer}, a Perl script that produces the final MetaScope report {\tt output.xml} in XML format.

The analyzer uses three criteria to decide which alignments are deemed significant and all non-significant
alignments are ignored in all subsequent analysis steps.
The first criterion is a minimum alignment bit score (option {\tt minscore}, default is 50).
Second, for each read we only consider alignments that have maximal bit score, or that are within $x\%$ of the
top score, where $x$ is set by a user option called  ${\tt top}$.

The third criterion, which is only applied  to Illumina reads, aims at ensuring that a significant alignment
covers a large proportion of the corresponding read.
Because the quality of an Illumina read tends to degrade toward the end of the read,
we  calculate the proportion of read covered as alignment length divided by
 ``covered prefix length'',
where the latter is the length of the prefix of the read up to the last base that is covered by the alignment.
In more detail, an alignment must fulfill   $1-P_{s}/P_{e}<\mbox{\tt minover}$  to be deemed
significant, where $P_{s}$ and $P_{e}$ are the alignment start and end position on the read and {\tt minover}
is a user-specified parameter (default is $0.8$).

\paragraph{Weighted LCA}

The assignment of reads to taxa based on a set of alignments to a reference database is often performed using the naive
LCA algorithm\cite{MEGAN2007} in which a read is placed on the lowest-common ancestor  of all
taxa in the NCBI taxonomy for which the read has a high-scoring alignment to a corresponding sequence in the reference database.
This approach is quite conservative and does not work well when there are multiple closely
related references in the database, as these  will move the assignment
to higher level on the phylogenetic tree. 

To overcome this, MetaScope uses a new weighted LCA  algorithm that proceeds in two rounds.
In the first round, the naive LCA is applied to all reads.
During this process, each reference sequence is assigned a weight that is the number of reads
that have a significant alignment to that reference sequence and for which the naive LCA assigns
the read to the same species that the reference sequence has.
Reference sequences that are not assigned a weight in this way are assigned weight~$1$.

In the second round, each read is then assigned to the lowest taxonomic node that lies above
a  fixed proportion (user parameter {\tt lca} default value $0.8$) of the sum of weights of reference sequences to which the read has a significant alignment. The lowest rank that we consider here is that of species. Reads that are assigned
to a sub-species or strain are moved up to the species level.

To address the problem of over-aggressive taxa assignment, for each assigned taxa node, we 
calculate and report the average alignment identity between the reads assigned to this node and the reference sequences.
If the average identity is below 90\% for a species level taxa node, a {\tt low\_identity} tag is  reported in the XML output
to indicate that a species-level assignment might be too aggressive.

\paragraph{Strain level assignment}
Our implementation of the weighted LCA assigns reads down to the level of species, but not further.
If the user requests strain-level analysis (option {\tt strain}) then the {\tt analyzer} proceeds as follows.
For each read that is assigned to a species node, we consider all alignments whose bit score are within $x$ percent
of the best score for the read, where $x$ is determined by a user parameter  {\tt strain\_top} (default value 10\%).
If a significant proportion (controlled by a user parameter {\tt strain\_lca}, default 80\%) of the best alignments agree on a strain
and these alignments have high sequence identity (controlled by parameter {\tt strain\_iden}, default 95\%), then the read is tentatively assigned to that strain. A strain is reported, if a significant proportion
(controlled by parameter {\tt strain\_report}, default 80\%) of the reads previously assigned to the species are tentatively assigned
to the strain.

\paragraph{Gene prediction}
To decide which genes to report for a given read, MetaScope produces two separate lists
of all genes that are partially covered by an alignment of the read. The first list is ranked by descending weight of reference sequence (as described above) and the second is ranked by descending coverage of genes (that is, by the number of
bases of the gene covered by any significant  alignment of any read). By default, Metascope reports the top five genes (user parameter
{\tt maxgene})  from each of the two ranked lists.

\subsection{Supporting data generation}

\paragraph{Data source}

All supporting data for MetaScope were downloaded from NCBI. The URLs for the data source are as follows:
\begin{itemize}
\item Human genome data: \url{ftp://ftp.ncbi.nlm.nih.gov/genomes/H_sapiens/}
\item GenBank data in ASN.1 format: \url{ftp://ftp.ncbi.nih.gov/ncbi-asn1/}
\item GenBank data in Genbank format: \url{ftp://ftp.ncbi.nih.gov/genbank/}
\item Taxonomy data: \url{ftp://ftp.ncbi.nlm.nih.gov/pub/taxonomy/}
\end{itemize}

\paragraph{Masking Human-like reference sequences}

Because virus and synthetic constructs often contain human sequences,
we decided to mask all human-like regions in those sequences in our working version of
the GenBank microbial database. To do this, we built a SASS index for the human genome reference
 in sensitive mode (SASS option {\tt --index-mode 2}). All virus and synthetic construct sequences were then shredded into 100bp fragments
with 50bp overlap. We used SASS to align the shredded sequences against
the human reference. If a shredded sequence aligned to the human genome with an alignment of 50 bases or more, and
at least 80\% identity, then the source region of the
shredded sequence covered by the alignment was masked by replacing all nucleotides by N's.

\paragraph{Data preprocessing}

DNA sequences were first extracted from GenBank ASN.1 data. Only sequences
under GenBank BCT, VRL, PHG, and SYN sections were included. The four
sections cover all GenBank sequences from bacteria, archaea, virus,
phage, and synthetic constructs. 

A mapping of GI numbers to NCBI taxon identifiers was extracted
from the GenBank ASN.1 files. We also extracted  taxonomic lineage information for each reference sequence
 from these files, rather than from
  the NCBI taxonomy dump file because only the ASN.1 files contain the correctly labeled strain description of reference sequences.
  The NCBI taxonomy file was used to complement the ASN.1-derived taxonomy data.

Information on protein coding regions, such as location, protein accession
number, locus tag, description, was extracted from the GenBank flat files. 

The set of scripts used to download process all reference data is
distributed in the \texttt{aux} folder of the MetaScope package.

\subsection{Parameters}
  
The output of different sequencing platforms varies in three main
aspects, namely the number of reads produced, the average read length
and the sequence quality. Individual Illumina datasets usually consist of
millions of reads with a read length of hundreds of base pairs.
Roche-454 datasets usually have less than one million reads, with a read length
 approaching 1000 bp. 
 Ion Torrent datasets contain hundreds of thousands
of reads, hundreds of base pairs long, with a lower level
of quality than the afore mentioned datasets. Finally, PacBio datasets are usually
smaller yet, with read lengths of thousands of base pairs, with a very high level of errors. 

\begin{table}
\centering{}%
\begin{tabular}{cccc}

Platform & SASS Index & {\tt top} & {\tt minover}\\
\hline 
Illumina & normal & 0.05 & 0.9\\
Ion Torrent & sensitive & 0.05 & 0\\
Roche 454 & normal & 0.05 & 0\\
PacBio & sensitive & 0.1 & 0\\
Unspecified & normal & 0.05 & 0\\
\end{tabular}
\caption{Default platform-specific settings used by MetaScope.}\label{settings-table}
\end{table}

To address these differences, MetaScope uses slightly different parameter
settings depending on which sequencing platform was used to generate
the input (see Supplement Table~\ref{settings-table}). For datasets with higher error
rates and smaller size, the pipeline uses SASS' sensitive
Genbank index so as to improve the detection of significant alignments
in the presence of sequencing errors. Moreover, in the taxonomic analysis
of such data, the pipeline employs a relaxed LCA with a {\tt top}
setting of 10\% so as to help avoid unreliable placement of reads
for PacBio data, but 5\% for other platforms. Because Illumina datasets  usually contain
millions of reads, here a even small sequence error rate can lead to a
large number of wrongly assigned reads. Hence, for Illumina we use a {\tt minover}
setting of 0.9 to ensure that  significant alignments cover at least 90\% of the high quality end of a read. 

The MetaScope parameters employed in the DTRA challenge differ slightly from the  default settings
recommended in Table~\ref{settings-table} due to the specific nature of the  DTRA testing datasets. 
Their metagenomic reads appeared to have originated from organisms whose genome sequences are well represented in Genbank
and thus they usually have a top-scoring alignment to the correct species (but also to many others).
In this situation, we were able to set {\tt top} to $0$ for all sequencing technologies except for PacBio, were $0.1$ was used to accommodate for the high rate of sequencing errors in PacBio data.

The default value for the {\tt maxgene} parameter (that controls the number of genes reported per read) is $5$, as this value
works well on all  DTRA challenge datasets. However, for the DTRA challenge Roche-454 datasets we used a value
of $1$ so has to achieve a particularly high gene score so as to compensate for low organisms scores on the Roche-454
test datasets.

\section*{Acknowledgements}

This research is partially supported by the National Research Foundation and Ministry of
Education Singapore under its Research Centre of Excellence Programme.

\section*{Author contributions}

All authors contributed equally to the development and implementation of the described software.

\section*{Bibliography}

\bibliographystyle{naturemag} {\footnotesize
\bibliography{compbio-2012}

\begin{thebibliography}{10}
\expandafter\ifx\csname url\endcsname\relax
  \def\url#1{\texttt{#1}}\fi
\expandafter\ifx\csname urlprefix\endcsname\relax\def\urlprefix{URL }\fi
\providecommand{\bibinfo}[2]{#2}
\providecommand{\eprint}[2][]{\url{#2}}

\bibitem{Handelsman2004}
\bibinfo{author}{Handelsman, J.}
\newblock \bibinfo{title}{Metagenomics: Application of genomics to uncultured
  microorganisms}.
\newblock \emph{\bibinfo{journal}{Microbiology and Molecular Biology Reviews}}
  \textbf{\bibinfo{volume}{68}}, \bibinfo{pages}{669--685}
  (\bibinfo{year}{2004}).

\bibitem{HumanMicrobiomeProject2007}
\bibinfo{author}{Turnbaugh, P.~J.} \emph{et~al.}
\newblock \bibinfo{title}{The {Human Microbiome Project}}.
\newblock \emph{\bibinfo{journal}{Nature}} \textbf{\bibinfo{volume}{449}},
  \bibinfo{pages}{804--810} (\bibinfo{year}{2007}).

\bibitem{Rusch2007}
\bibinfo{author}{Rusch, D.~B.} \emph{et~al.}
\newblock \bibinfo{title}{The {Sorcerer II Global Ocean Sampling} expedition:
  northwest {Atlantic} through eastern tropical {Pacific}}.
\newblock \emph{\bibinfo{journal}{PLoS Biol}} \textbf{\bibinfo{volume}{5}},
  \bibinfo{pages}{e77} (\bibinfo{year}{2007}).
\newblock \urlprefix\url{http://dx.doi.org/10.1371/journal.pbio.0050077}.

\bibitem{Nielsen2011}
\bibinfo{author}{Albertsen, M.}, \bibinfo{author}{Hansen, L.~B.},
  \bibinfo{author}{Saunders, A.~M.}, \bibinfo{author}{Nielsen, P.~H.} \&
  \bibinfo{author}{Nielsen, K.~L.}
\newblock \bibinfo{title}{A metagenome of a full-scale microbial community
  carrying out enhanced biological phosphorus removal}.
\newblock \emph{\bibinfo{journal}{ISME J}}  (\bibinfo{year}{2011}).
\newblock \urlprefix\url{http://dx.doi.org/10.1038/ismej.2011.176}.

\bibitem{Mackelprang2011}
\bibinfo{author}{Mackelprang, R.} \emph{et~al.}
\newblock \bibinfo{title}{Metagenomic analysis of a permafrost microbial
  community reveals a rapid response to thaw}.
\newblock \emph{\bibinfo{journal}{Nature}} \textbf{\bibinfo{volume}{480}},
  \bibinfo{pages}{368--371} (\bibinfo{year}{2011}).

\bibitem{Schuenemann2011}
\bibinfo{author}{Schuenemann, V.~J.} \emph{et~al.}
\newblock \bibinfo{title}{Targeted enrichment of ancient pathogens yielding the
  ppcp1 plasmid of yersinia pestis from victims of the black death}.
\newblock \emph{\bibinfo{journal}{Proceedings of the National Academy of
  Sciences}} \textbf{\bibinfo{volume}{108}}, \bibinfo{pages}{E746--E752}
  (\bibinfo{year}{2011}).
\newblock \urlprefix\url{http://www.pnas.org/content/108/38/E746.abstract}.
\newblock \eprint{http://www.pnas.org/content/108/38/E746.full.pdf+html}.

\bibitem{GEBA2009}
\bibinfo{author}{Wu, D.} \emph{et~al.}
\newblock \bibinfo{title}{A phylogeny-driven genomic encyclopaedia of bacteria
  and archaea}.
\newblock \emph{\bibinfo{journal}{Nature}} \textbf{\bibinfo{volume}{462}},
  \bibinfo{pages}{1056--1060} (\bibinfo{year}{2009}).

\bibitem{Wheeler2008}
\bibinfo{author}{Wheeler, D.~L.} \emph{et~al.}
\newblock \bibinfo{title}{Database resources of the {National Center for
  Biotechnology Information}}.
\newblock \emph{\bibinfo{journal}{Nucleic Acids Res}}
  \textbf{\bibinfo{volume}{36}}, \bibinfo{pages}{D13--D21}
  (\bibinfo{year}{2008}).
\newblock \urlprefix\url{http://dx.doi.org/10.1093/nar/gkm1000}.

\bibitem{MEGAN2007}
\bibinfo{author}{Huson, D.~H.}, \bibinfo{author}{Auch, A.~F.},
  \bibinfo{author}{Qi, J.} \& \bibinfo{author}{Schuster, S.~C.}
\newblock \bibinfo{title}{{MEGAN} analysis of metagenomic data}.
\newblock \emph{\bibinfo{journal}{Genome Res}} \textbf{\bibinfo{volume}{17}},
  \bibinfo{pages}{377--386} (\bibinfo{year}{2007}).
\newblock \urlprefix\url{http://dx.doi.org/10.1101/gr.5969107}.

\bibitem{MEGAN2011}
\bibinfo{author}{Huson, D.~H.}, \bibinfo{author}{Mitra, S.},
  \bibinfo{author}{Weber, N.}, \bibinfo{author}{Ruscheweyh, H.-J.} \&
  \bibinfo{author}{Schuster, S.~C.}
\newblock \bibinfo{title}{Integrative analysis of environmental sequences using
  {MEGAN\,4}}.
\newblock \emph{\bibinfo{journal}{Genome Research}}
  \textbf{\bibinfo{volume}{21}}, \bibinfo{pages}{1552--1560}
  (\bibinfo{year}{2011}).

\bibitem{altschul90}
\bibinfo{author}{Altschul, S.~F.}, \bibinfo{author}{Gish, W.},
  \bibinfo{author}{Miller, W.}, \bibinfo{author}{Myers, E.~W.} \&
  \bibinfo{author}{Lipman, D.~J.}
\newblock \bibinfo{title}{Basic local alignment search tool}.
\newblock \emph{\bibinfo{journal}{Journal of Molecular Biology}}
  \textbf{\bibinfo{volume}{215}}, \bibinfo{pages}{403--410}
  (\bibinfo{year}{1990}).

\bibitem{Burkhardt01}
\bibinfo{author}{Burkhardt, S.} \& \bibinfo{author}{K{\"a}rkk{\"a}inen, J.}
\newblock \bibinfo{title}{Better filtering with gapped q-grams}.
\newblock \emph{\bibinfo{journal}{Fundamenta Informaticae}}
  \textbf{\bibinfo{volume}{XXIII}}, \bibinfo{pages}{1001--1018}
  (\bibinfo{year}{2001}).

\bibitem{Ma02}
\bibinfo{author}{Ma, B.}, \bibinfo{author}{Tromp, J.} \& \bibinfo{author}{Li,
  M.}
\newblock \bibinfo{title}{{PatternHunter}: faster and more sensitive homology
  search}.
\newblock \emph{\bibinfo{journal}{Bioinformatics}}
  \textbf{\bibinfo{volume}{18}}, \bibinfo{pages}{440--445}
  (\bibinfo{year}{2002}).

\bibitem{SeqAn}
\bibinfo{author}{D\"oring, A.}, \bibinfo{author}{Weese, D.},
  \bibinfo{author}{Rausch, T.} \& \bibinfo{author}{Reinert, K.}
\newblock \bibinfo{title}{{SeqAn} an efficient, generic {C++} library for
  sequence analysis}.
\newblock \emph{\bibinfo{journal}{BMC Bioinformatics}}
  \textbf{\bibinfo{volume}{9}}, \bibinfo{pages}{11--11} (\bibinfo{year}{2008}).

\bibitem{Myers1999b}
\bibinfo{author}{Myers, E.}
\newblock \bibinfo{title}{A fast bit-vector algorithm for approximate string
  matching based on dynamic programming}.
\newblock \emph{\bibinfo{journal}{Journal of the ACM}}
  \textbf{\bibinfo{volume}{46}}, \bibinfo{pages}{395--415}
  (\bibinfo{year}{1999}).

\bibitem{Margulies2005short}
\bibinfo{author}{Margulies, M.} \& \bibinfo{author}{{\it et al.}}
\newblock \bibinfo{title}{Genome sequencing in microfabricated high-density
  picolitre reactors}.
\newblock \emph{\bibinfo{journal}{Nature}} \textbf{\bibinfo{volume}{437}},
  \bibinfo{pages}{376--380} (\bibinfo{year}{2005}).

\bibitem{boost_graph}
\bibinfo{author}{Siek, J.}, \bibinfo{author}{Lee, L.-Q.} \&
  \bibinfo{author}{Lumsdaine, A.}
\newblock \bibinfo{title}{Boost random number library}.
\newblock \bibinfo{howpublished}{http://www.boost.org/libs/graph/}
  (\bibinfo{year}{2000}).

\bibitem{Bowtie2}
\bibinfo{author}{Langmead, B.} \& \bibinfo{author}{Salzberg, S.}
\newblock \bibinfo{title}{Fast gapped-read alignment with {Bowtie\,2}}.
\newblock \emph{\bibinfo{journal}{Nat Meth}} \textbf{\bibinfo{volume}{9}},
  \bibinfo{pages}{357--359} (\bibinfo{year}{2012}).

\bibitem{BWA}
\bibinfo{author}{Li, H.} \& \bibinfo{author}{Durbin, R.}
\newblock \bibinfo{title}{Fast and accurate short read alignment with
  {Burrows-Wheeler} transform.}
\newblock \emph{\bibinfo{journal}{Bioinformatics (Oxford, England)}}
  \textbf{\bibinfo{volume}{25}}, \bibinfo{pages}{1754--1760}
  (\bibinfo{year}{2009}).

\bibitem{Ilie:2011fk}
\bibinfo{author}{Ilie, L.}, \bibinfo{author}{Ilie, S.},
  \bibinfo{author}{Khoshraftar, S.} \& \bibinfo{author}{Bigvand, A.~M.}
\newblock \bibinfo{title}{Seeds for effective oligonucleotide design}.
\newblock \emph{\bibinfo{journal}{BMC Genomics}} \textbf{\bibinfo{volume}{12}},
  \bibinfo{pages}{280} (\bibinfo{year}{2011}).

\bibitem{LocalAlign}
\bibinfo{author}{Smith, T.~F.} \& \bibinfo{author}{Waterman, M.~S.}
\newblock \bibinfo{title}{Identification of common molecular subsequences}.
\newblock \emph{\bibinfo{journal}{JMB}} \textbf{\bibinfo{volume}{147}},
  \bibinfo{pages}{195--197} (\bibinfo{year}{1981}).

\end{thebibliography}
}

\end{document}